\input{aipcheck.tex}
\documentclass[final]{aipproc}
\layoutstyle{8x11double}

\SetInternalRegister\hbadness{8000}

\begin{document}

\title{Evolution of X-ray spectra of Cygnus X-3 with radio flares}

\author{Manojendu Choudhury}{address={Tata Institute of Fundamental Research, Mumbai -- 400005. INDIA}}
\author{A. R. Rao}{address={Tata Institute of Fundamental Research, Mumbai -- 400005. INDIA}}

\begin{abstract}
Cygnus X-3, among the X-ray binaries, is one of the brightest in the radio band,
repeatedly exhibiting huge radio flares. The X-ray spectra shows two definite states,
low (correspondingly hard) and high (correspondingly soft). During the hard state the
X-ray spectra shows a pivoting behaviour correlated to the radio emission. In the
high state the X-ray spectra shows a gamut of behaviour which controls the radio
flaring activity of the source. The complete evolution of the X-ray spectra along
with the radio flaring activity is reported here, for the first time for this source.

\end{abstract}

\maketitle

\section{Introduction}
Cygnus X-3 is one of the most studied yet least understood source among the X-ray
binaries. The nature of the compact object is still unresolved, although the X-ray
spectral energy distribution (SED) suggests it to be a black hole candidate. Also,
similar to the established black hole candidates, it shows two main states of X-ray
emission, high (correspondingly 'soft') and low (correspondingly 'hard'), although
these states can not be classified as the canonical black hole states as the
individual spectral components during these states are not those pertaining to the
classical black hole candidates, viz. Cygnus X-1 \cite{cho03}. Cygnus X-3 is one of
the brightest and persistant radio source among the X-ray binaries, exhibiting huge
radio flares quite frequently. These flares occur during the X-ray high (and
correspondingly soft) state. 

\subsection{General X-ray spectra}
The X-ray SED of this source typically shows two different states, low
(correspondingly hard) and high (correspondingly soft), see Figure 1. There is very
high intrinsic absorption of the X-ray emission in this system, probably due to the
dust originating from the winds of the Wolf-Rayet type companion star. As a result
the disk black body component is not observable using the \emph{RXTE-PCA} during the
low state, and the spectra is best described by a comptonising component, CompST, and
a power law. The high state is generally characterised by a thermal component,
multi- coloured disk black body, plus a comptonising component (CompST)
\cite{cho02b}, except on a few cases when the disk black body component is replaced
by a power law, anologous to the low state (see the following sections). Spectral
analysis of the data obtained from ASCA observatory also reveal three Fe emission
lines at 6.36, 6.67 \& 6.96 keV. These line features are more prominent during the
low state (Figure 1).

\begin{figure}
\includegraphics[height=.25\textheight]{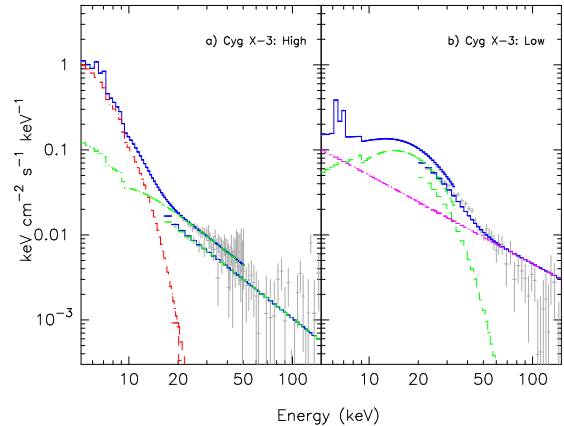}
\caption{Typical X-ray spectral energy distribution of Cygnus X-3 during its high
(correspondingly soft) and low (correspondingly hard) states, along with their
continuum components.The high state is characterised by disk black body component
and a comptonising component, (CompST). The low state is charecterised by CompST and
a power law.}
\end{figure}

\section{Long term monitoring in X-ray and radio bands}
The \emph{All Sky Monitor (ASM)} aboard the \emph{RXTE} satellite observatory, the
\emph{Burst And Transient Sources Experiment (BATSE)} aboard the \emph{CGRO}
satellite observatory, and the \emph{Green Bank Interferometer (GBI)}
serendipitously monitored the source (quasi-)simultaneously in the soft X-ray
(2-12 keV), hard X-ray (20-100 keV) and the radio (2.2 GHz) band, respectively,
during the period MJD 50400 -- 51500. Thereafter \emph{CGRO-BATSE} stopped its
operation while \emph{GBI} and \emph{RXTE-ASM} continued the simultaneous monitoring
until \emph{GBI} stopped its operation around MJD 51700. 

\begin{figure}
\includegraphics[height=.25\textheight]{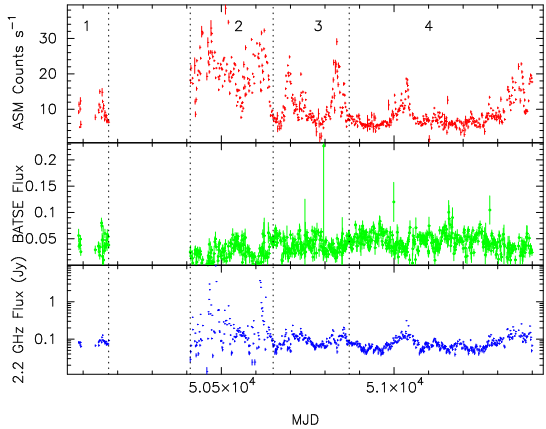}
\caption{Soft X-ray (\emph{RXTE-ASM}, 2-12 keV), hard X-ray (\emph{CGRO-BATSE},
20-100 keV) \& radio (\emph{GBI}, 2.2 GHz) monitoring of the source. Region 1, 3 \& 4
correspond to the low state whereas region 2 correspond to the high state of the
X-ray emission.}
\end{figure}

Figure 2 shows the light curves of the source as observed by the three observatories.
The regions 1, 3 \& 4 correspond to the hard (as well as `low') state of the X-ray
emission, with correlated radio emission \cite{cho02a, cho03}. The region 2
correspond to the high (correspondingly `soft') state of the X-ray emission. Within
the precincts of the hard (low) state of the X-ray emission, the soft X-ray
(\emph{RXTE-ASM}) and the radio (\emph{GBI}) are highly correlated, while the hard
X-ray (\emph{CGRO-BATSE}) is anti-correlated to both the soft X-ray and the radio.
This is due to the pivoting in the X-ray spectra, correlated to the radio emission.
In the high (soft) state, we report the evidence of ejection of the central
non-thermal Compton cloud, giving rise to very bright radio flares. 

\begin{figure}[b]
\includegraphics[height=.17\textheight]{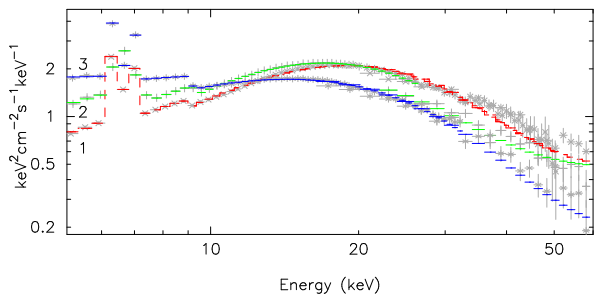}
\caption{Pivoting of the X-ray spectra, correlated to the radio emission, in the low
(correspondingly hard) state. The spectra becomes softer with increased radio
emission.}
\end{figure}

\section{Low (hard) state: Pivoting of non-thermal spectra}

The low (as well as hard) state spectral energy distribution of the X-ray emission is
generally non-thermal in nature (Figure 1). Figure 3 shows the pivoting of the X-ray
spectra, which is correlated to the radio emission, in this state. The soft X-ray
flux increases with the radio emission, inversely the hard X-ray flux spectra hardens
with decrease in radio flux, with the pivot point lying in the region $10-20$ keV
\cite{cho02a, cho03}.

Assuming that the region of the Comptonization is confined to a small volume near the
compact obejct, the pivoting of the spectra can be qualitatively explained by the Two
Component Accretion Flow model of \cite{cha96}. The Comptonisation component of the
spectra originates from a region close to the compact object, confined within the
Centrifugal Boundary Layer (CENBOL). At low accretion rate, the CENBOL is far away
from the compact object and the spctrum is harder with lower outflow \cite{das99}.
On increasing the accretion rate the CENBOL comes closer to the compact object with
greater outflow, giving rise to increased radio emission and decreased non-thermal
hard X-ray emission. The radio emission, from a core jet, is inversely proportional
to the compression ratio which decreases as the spectra softens, in the low (hard)
state. 

\section{High (soft) state: X-ray spectral evolution drives the radio emission}

\begin{figure}
\includegraphics[height=.6\textheight, angle=-90]{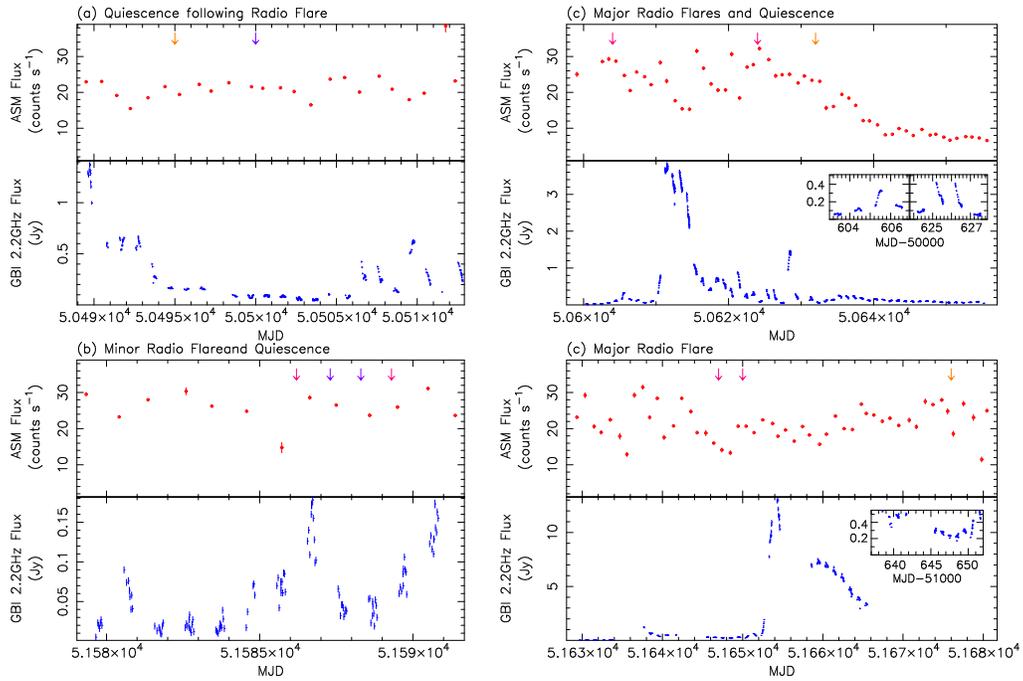}
\caption{The soft X-ray (\emph{RXTE-ASM}, 2-12 keV) and radio (\emph{GBI}, 2.2GHz)
monitoring of the source during the X-ray high state. The days for which the X-ray
spectra obtained from the pointed observations using \emph{RXTE-PCA} are reported
here are indicated by arrows. The insets in the two right hand panels highlight the
minor flares.}
\end{figure}

\begin{figure}
\includegraphics[height=.5\textheight, angle=-90]{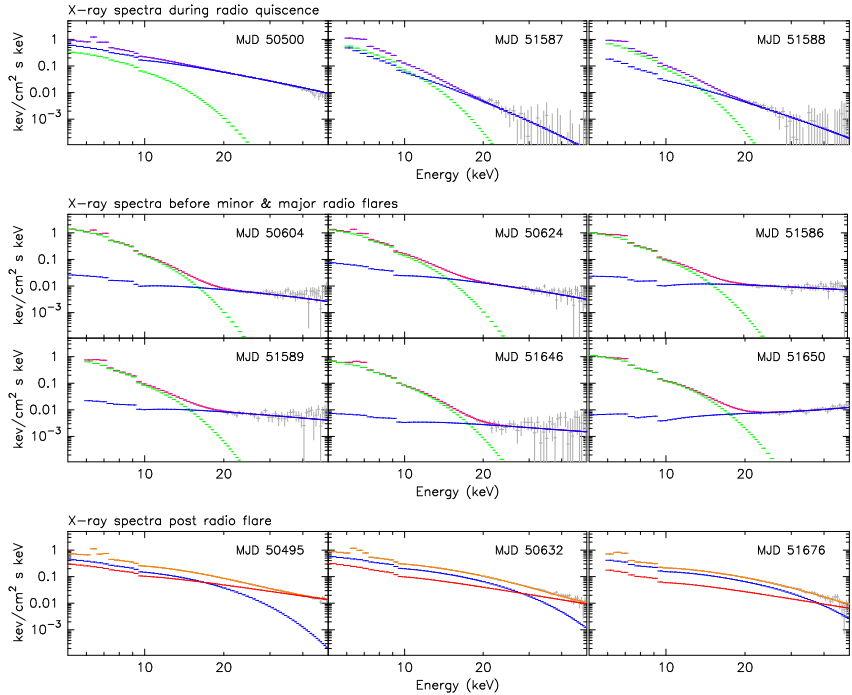}
\caption{The X-ray spectral energy distribution (SED) and the individual continuum components, during the radio quiescent, pre-radio flare \& post-radio flare phases. The quiescent phase has disk black body and Comptonising component at near equal ratio, the pre-radio flare has vanishingly small Comptonising component, and the post-radio flare has the disk black body component replaced by a simple power law.}
\end{figure}

\begin{table}
\begin{tabular}{lccccccccc}
\hline
\tablehead{3}{l}{b}{Quiescent Radio Emission} & & & & & & & \\
& & \tablehead{3}{c}{}{Disk Black Body} & & \tablehead{4}{c}{}{CompST} \\
\cline{3-5} 
\cline{7-10}
\tablehead{1}{l}{}{MJD} & \tablehead{1}{c}{}{Total Flux} & \tablehead{1}{c}{}{$kT_{in}$ (keV)} & \tablehead{1}{c}{}{Flux} & \tablehead{1}{c}{}{\% of Total Flux} & & \tablehead{1}{c}{}{$kT_E$ (keV)} & \tablehead{1}{c}{}{$\tau$} & \tablehead{1}{c}{}{Flux} & \tablehead{1}{c}{}{\% of Total Flux} \\
\hline
50500 & 7.9 & 1.8 & 1.9 & 24.05 & & 23.93 & 2.22 & 6.0 & 75.95 \\
51587 & 6.6 & 1.44 & 3.4 & 51.51 & & 13.39 & 1.59 & 3.2 & 48.49 \\
51588 & 5.1 & 1.42 & 3.7 & 72.55 & & 21.69 & 1.39 & 1.4 & 27.45 \\
\hline
\tablehead{3}{l}{b}{Pre-Radio Flare} & & & & & & & \\
& & \tablehead{3}{c}{}{Disk Black Body} & & \tablehead{4}{c}{}{CompST} \\
\cline{3-5} 
\cline{7-10}
\tablehead{1}{l}{}{MJD} & \tablehead{1}{c}{}{Total Flux} & \tablehead{1}{c}{}{$kT_{in}$ (keV)} & \tablehead{1}{c}{}{Flux} & \tablehead{1}{c}{}{\% of Total Flux} & & \tablehead{1}{c}{}{$kT_E$ (keV)} & \tablehead{1}{c}{}{$\tau$} & \tablehead{1}{c}{}{Flux} & \tablehead{1}{c}{}{\% of Total Flux} \\
\hline
50604 & 6.6 & 1.53 & 6.0 & 90.91 & & 18.27 & 4.07 & 0.6 & 9.09 \\
50624 & 6.9 & 1.55 & 5.9 & 85.51 & & 18.34 & 3.31 & 1.0 & 14.49 \\
51586 & 5.3 & 1.56 & 4.4 & 83.02 & & 42.31 & 3.41 & 0.9 & 16.98 \\
51589 & 4.5 & 1.53 & 3.8 & 84.44 & & 53.53 & 2.40 & 0.7 & 15.56 \\
51646 & 3.6 & 1.63 & 3.3 & 91.67 & & 54.33 & 2.53 & 0.3 & 8.33 \\
51650 & 5.8 & 1.70 & 5.0 & 86.21 & & 80.79 & 9.91 & 0.8 & 13.79 \\
\hline
\tablehead{3}{l}{b}{Post-Radio Flare} & & & & & & & \\
& & \tablehead{3}{c}{}{Power law} & & \tablehead{4}{c}{}{CompST} \\
\cline{3-5} 
\cline{7-10}
\tablehead{1}{l}{}{MJD} & \tablehead{1}{c}{}{Total Flux} & \tablehead{1}{c}{}{$\Gamma$} & \tablehead{1}{c}{}{Flux} & \tablehead{1}{c}{}{\% of Total Flux} & & \tablehead{1}{c}{}{$kT_E$ (keV)} & \tablehead{1}{c}{}{$\tau$} & \tablehead{1}{c}{}{Flux} & \tablehead{1}{c}{}{\% of Total Flux} \\
\hline
50495 & 8.5 & 2.43 & 4.2 & 49.41 & & 4.03 & 8.12 & 4.3 & 50.59 \\
50632 & 9.9 & 2.62 & 4.0 & 40.40 & & 5.12 & 7.05 & 5.9 & 59.60 \\
51676 & 7.9 & 2.62 & 2.7 & 34.18 & & 6.18 & 6.36 & 5.2 & 65.82 \\
\hline \hline
\end{tabular}
\caption{Model parameters of the continuum components and their flux contributions of
the X-ray SED}
\label{tab:a}
\end{table}

The soft X-ray (\emph{RXTE-ASM}, 2 -- 12 keV) and the radio (\emph{GBI}, 2.2 GHz)
monitoring shows a more complicated evolution of the high energy emission with
respect to the emission in the radio band (Figure 4), in the high (correspondingly
soft) state of X-ray emission. The X-ray spectra in this state is generally dominated
by the thermal multicoloured disk black body component, along with a hard component
best described by a CompST model \cite{sun80}, except for the post flare phase, when
the spectral shape hardens in the soft X-ray region. As shown in Figure 5, the
salient features of the spectra are:-

\begin{description}
\item {\bf The radio quiescent phase.} The X-ray spectra has a strong disk black body
and an equally strong Comptonising component.
\item {\bf Pre-radio flare.} The Comptonising component vanishes, resulting in a
flare. The flare may result in a time scale of a day or less.
\item {\bf Post-radio flare.} The succession of radio flares, both minor as well as
major, is stopped by the change in the X-ray spectrum, with the spectral shape
hardening in the soft X-ray region. During this phase the disk black body component
becomes insignificant, the spectral shape is explained by the similar combination of
non-thermal components, viz. CompST and power law.
\end{description}

The evolution of the X-ray spectral changes with respect to the radio emission can be explained as follows:-
\begin{enumerate}
\item The radio quiescent emission is marked by the radio emission (2.2 GHz)
bordering around 110 mJy and below. The X-ray spectra has strong CompST component
with the amounting $75\% - 25\%$ of the total flux.
\item The vanishing of the CompST component (flux going below 15\%) always precedes
a minor flare, with the radio flux around $\approx150 - 800$ mJy, suggesting the
ejection of the central Compton cloud resulting in the flare. The stronger the
ejection, the louder the flare.
\item The minor flare may be followed by the filling of the central Compton cloud,
i.e. increase in CompST flux, causing the radio emission to become quiescent.
Otherwise, if the continuous accretion persists with the central cloud unfilled, i.e.
the CompST flux remains low, a major radio flare (2.2 GHz, flux $> 1$ Jy) follows.
\item The continuing series of minor and major flares come to an end only with the
change in the X-ray spectra, i.e. hardening of the soft X-ray band, with the flux
level remaining high. This is the most interesting state of the X-ray spectra with
the shape being best fit by the model of the low (correspondingly hard) state, i.e.
power law and CompST, although the soft X-ray flus remains high. This change in the
X-ray spectra puts a brake in the episodes of radio flaring.
\end{enumerate}

\bibliographystyle{aipprocl}

\bibliography{choudhurym}

\end{document}